\documentclass[article, superscriptaddress, nofootinbib, showpacs,10pt]{revtex4}
\usepackage[titletoc,title,toc]{appendix}
\usepackage{graphicx}
\usepackage{amssymb}
\usepackage{epstopdf}
\usepackage{amsmath}
\usepackage{hyperref}
\usepackage{mathrsfs}
\usepackage{varioref}
\usepackage[active]{srcltx}
\usepackage{longtable}
\usepackage{amstext}
\usepackage{epsfig}
\usepackage{esint}
\expandafter\ifx\csname package@font\endcsname\relax\else
\expandafter\expandafter
\expandafter\usepackage
\expandafter\expandafter
\expandafter{\csname package@font\endcsname}%
\fi
\hypersetup{
    bookmarks=false,         % show bookmarks bar?
    unicode=false,          % non-Latin characters in Acrobat's bookmarks
    pdftoolbar=true,        % show Acrobat's toolbar?
    pdfmenubar=true,        % show Acrobat's menu?
    pdffitwindow=false,     % window fit to page when opened
    pdfstartview={FitH},    % fits the width of the page to the window
    pdftitle={My title},    % title
    pdfauthor={Author},     % author
    pdfsubject={Subject},   % subject of the document
    pdfcreator={Creator},   % creator of the document
    pdfproducer={Producer}, % producer of the document
    pdfkeywords={keyword1} {key2} {key3}, % list of keywords
    pdfnewwindow=true,      % links in new window
    colorlinks=true,       % false: boxed links; true: colored links
    linkcolor=red,          % color of internal links
    citecolor=cyan,        % color of links to bibliography
    filecolor=magenta,      % color of file links
    urlcolor=black,           % color of external links
    linktocpage=true
}
\DeclareFontFamily{OT1}{pzc}{}
\DeclareFontShape{OT1}{pzc}{m}{it}%
             {<-> s * [0.900] pzcmi7t}{}
\DeclareMathAlphabet{\mathscr}{OT1}{pzc}%
                                 {m}{it}
\labelformat{section}{Section #1} 
\labelformat{subsection}{Section #1} 
\labelformat{equation}{Eq.(#1)} 
\labelformat{figure}{Fig.~#1} 
\labelformat{subfigure}{Fig.~\thefigure#1} 
\labelformat{table}{Tab.~#1} 
\newcommand{\be}{\begin{equation}}
\newcommand{\ee}{\end{equation}}
\newcommand{\ba}{\begin{eqnarray}}
\newcommand{\ea}{\end{eqnarray}}
\newcommand{\ban}{\begin{eqnarray*}}
\newcommand{\ean}{\end{eqnarray*}}

\def\rt{\right}

\begin{document}

\title{Gravitational Lensing by Self-Dual Black
Holes in Loop Quantum Gravity}

\author{Satyabrata Sahu}
\email{satyabrata@tifr.res.in}
\affiliation{Tata Institute of Fundamental Research\\
Homi Bhabha Road, Mumbai 400005, India}
\author{Kinjalk Lochan}
\email{kinjalk@iucaa.ernet.in}
\affiliation{Tata Institute of Fundamental Research\\
Homi Bhabha Road, Mumbai 400005, India}
\affiliation{Inter-University Centre for Astronomy and Astrophysics\\ Ganeshkhind, Pune 411 007, India}
\author{D. Narasimha}
\email{dna@tifr.res.in}
 \affiliation{Tata Institute of Fundamental Research\\
 Homi Bhabha Road, Mumbai 400005, India}

 \begin{abstract}
We study gravitational lensing by a recently proposed black hole solution in Loop Quantum Gravity. We highlight the fact that the quantum gravity corrections to the Schwarzschild metric in this model evade the `'mass suppression'' effects (that the usual quantum gravity corrections are susceptible to) by virtue of one of the parameters in the model being dimensionless, which is unlike any other quantum gravity motivated parameter. Gravitational lensing in the strong and weak deflection regimes is studied and a sample consistency relation is presented which could serve as a test of this model. We discuss that though the consistency relation for this model is qualitatively similar to what would have been in Brans-Dicke, in general it can be a good discriminator between many alternative theories. Although the observational prospects do not seem to be very optimistic even for a galactic supermassive black hole case, time delay between relativistic images for billion solar mass black holes  in other galaxies might be within reach of future relativistic lensing observations. 

\end{abstract}
\maketitle

\section{Introduction}
The construction of a consistent theory of quantum gravity describing
our universe remains a cherished dream of theoretical physicists.
Two foremost candidates for a theory of quantum gravity at present are String Theory
and Loop Quantum Gravity (LQG) among many. There are multiple pointers to believe that study of
black holes is likely to yield a few insights for revealing the quantum
nature of gravity and hence the quantum nature of spacetime. The singularity
problem, the entropy puzzle and the information loss paradox (especially in light of some recent developments \cite{Braunstein:2009my,Almheiri:2012rt,Mathur:2009hf}) leads one to
believe that black hole might be fundamentally quantum object in some
sense. Motivated by these developments, it has been recently argued that precision interferometry of lensed pulsar images may shed light on the quantum gravitational processes near black holes \cite{Pen:2013qva}. Large enough black holes would be described by
a classical black hole metric which can encode the quantum corrections
in a suitable way. For example, in the spherically symmetric case, %(with which we will be concerned here)
the deviation from the Schwarzschild metric can be characterized by a parameter (apart from mass) 
involved in the corresponding quantum gravity theory such as the string tension in the string theory case or a non-commutative parameter in certain non-commutative quantum geometry scenarios or the Barbero-Immirzi parameter for LQG.
Of course one expects the quantum gravity effects to be relevant when spacetime curvature reaches the Planckian regime which is reflected by the fact that string
tension or non-commutativity parameter have dimension of $[L]^2$ and hence their effect
on observables is mass suppressed i.e. typically the observables are corrected by factors of $1+\left(\frac{l_{qg}}{M}\right)^{\mathscr{m}}$ where $l_{qg}$ is a quantum gravity scale and $\mathscr{m}$ is an arbitrary positive power. Hence the correction is important for very low mass black holes, indeed microscopic ones if $l_{qg}$ is near Planck scale $\ell_{P}$\footnote{There are suggestions that $l_{qg}$ could be far below the Planck scale because of the large entropy of black holes. The reason
 is that when a large number $N$ is involved, the length scale of
quantum gravitational effects need not be $\ell_{P}$ but $N^\mathscr{n} \ell_{P}$ for some $\mathscr{n} > 0$ and for black holes the large number concerned is the large number of black hole microstates.}.
Nonetheless it has been of some interest to study the effects of such
corrections both quantitatively and qualitatively for string inspired
scenarios \cite{Bhadra:2003zs} and non-commutative  black holes \cite{Ding:2010dc}. These studies have concentrated on gravitational lensing as an observational probe, especially in
the strong field regime. The strong field lensing coefficients are
interesting in their own right too, owing to their relationship with
quasinormal modes \cite{Stefanov:2010xz} and absorption cross sections from black holes \cite{Wei:2011zw}. In
this paper we study gravitational lensing in a loop quantum gravity inspired scenario \cite{modesto1}
both in the strong and weak field regimes. The peculiarity of the particular
solution we study, as we shall discuss in detail later, is that the
quantum gravity corrections, apart from entering into the metric function
as a dimension $L^2$ term viz. the minimal area quanta, also enter as a dimensionless parameter, viz., the `polymeric parameter'. This is in contrast to the stringy and non-commutative
 cases discussed before and makes the lensing observations evade
the mass suppression effect and is by far the most observationally
favorable quantum gravity scenario, though still quite difficult observational
challenge in itself.

The paper is organized as follows. In section \ref{lqbh} the loop black hole spacetime is introduced and in sections \ref{weak} and \ref{strong} respectively, the weak field and strong field lensing signatures are studied. After a brief discussion of observational prospects in section \ref{obs} and the contrast with other quantum gravity scenarios in section \ref{cont}, a sample consistency relation between the far field and relativistic Einstein Rings is presented in section \ref{erc}. Finally we conclude with discussions in section \ref{diss}.
\section{A Black Hole Solution in LQG}\label{lqbh} 
Let us briefly discuss the black hole solution in LQG.
LQG is a potential candidate theory of quantum gravity \cite{Loop,Bilson-Thompson:2014hoa}. In this theory, ${ su}(2)$ valued connection  (evaluated for the holonomies along closed loops) and triad densities serve as
the canonical variables for quantization.
The solution of this theory is obtained in terms of constraint equations (Gauss, Diffeomorphism and Hamiltonian constraints respectively)
whose solutions will describe the dynamics of space time. Though the theory is well understood at the kinematical level in terms
of solutions of Gauss and diffeomorphism constraints known as `spin-networks', solution for all the constraints in the full generality
has yet not been achieved. 
The theory tentatively
suggests discrete structure of space-time at the kinematical level. Geometric quantities like area, volume are shown to have discrete 
spectra (for the Euclidean case which can be carried over to Lorentzian case at the kinematical level).
However, the full dynamical structure of the theory is missing
as the solving the Hamiltonian constraint remains an open issue.
 
\subsection{Connection Formulation for Schwarzschild}
In Loop Quantum Gravity, black holes in different spacetimes have been studied \cite{modesto1} employing the semiclassical
techniques in the minisuperspace quantization scheme. In this work we consider a black hole model for the quantum corrected
Schwarzschild spacetime \cite{modesto2}, which is characterized by a `polymeric parameter' $\delta$,
which is a measure of `quantum correction' to the classical description. In order to study the loop corrected behavior of 
Schwarzschild black hole, one starts with the homogeneous but anisotropic Kantowski-Sachs spacetime expressed in terms of 
Ashtekar variables. 
\begin{eqnarray}
&& {\cal A}= \tilde{c} \tau_3 d x + \tilde{b} \tau_2 d \theta - \tilde{b} \tau_1 \sin \theta d \phi + \tau_3 \cos \theta d \phi,
\nonumber \\
&&E = \tilde{p}_c \tau_3 \sin \theta \frac{\partial}{\partial x} + \tilde{p}_b \tau_2 \sin \theta \frac{\partial}{\partial \theta} - \tilde{p}_b \tau_1 \frac{\partial}{\partial \phi}.
\label{contriad}
\end{eqnarray}
with $\tau_i\in {su}(2)$. Therefore, the basic canonical variables in this approach are $\{\tilde{b},\tilde{c};\tilde{p}_b,\tilde{p}_c\}$.
The Hamiltonian constraint in terms of these variables is given as
\begin{eqnarray}
\mathcal{C}_{H} = - \int \frac{N d x \sin \theta d \theta d \phi}{8 \pi  G_N \gamma^2}
\left[ (\tilde{b}^2 + \gamma^2) \frac{\tilde{p}_b \, {\rm sgn} (\tilde{p}_c)}{\sqrt{|\tilde{p}_c|}} + 2 \tilde{b} \tilde{c} \, \sqrt{|\tilde{p}_c|} \right], %d x \sin \theta d \theta d \phi.
\label{Ham1}
\end{eqnarray}
with $N$ giving the lapse function on the spatial surface given by the the metric $q^{ab}$ related to the triad $E^a_i$ as
$$E^a_i E^b_j \delta^{i j}= {\rm det}(q) q^{ab},$$
and $\gamma$ being the ``Barbero-Immirzi'' parameter which appears in the action of the theory, which at the classical level, is an equivalent canonical formulation of Einstein-Hilbert action. This parameter remains free in the quantization scheme and characterizes different unitarily inequivalent sectors in the quantum theory.

The integration over $x$ is restricted to a finite interval $L_0$ such that the Hamiltonian constraint appears simply as 
\begin{eqnarray}
\mathcal{C}_{H} = - \frac{N}{2 G_N \gamma^2} \left[ (b^2 + \gamma^2) \frac{p_b \, {\rm sgn} (p_c)}{\sqrt{|p_c|}} + 2 b c \, \sqrt{|p_c|} \right],
\label{Ham2}
\end{eqnarray}
in terms of rescaled variables given as
\begin{eqnarray}
b=\tilde{b}\hspace{0.5 in} c= L_0 \tilde{c}, \nonumber\\
p_b = L_0 \tilde{p}_b\hspace{0.5 in} p_c=\tilde{p}_c.
\end{eqnarray}
Further, fixing the gauge $N$ to $\gamma \, \sqrt{|p_c|} \, {\rm sgn} (p_c)/ b$, the Hamiltonian constraint can be further simplified to 
\begin{eqnarray}
\mathcal{C}_{H} = - \frac{1}{2 G_N \gamma} \left[ (b^2 + \gamma^2) p_b/b + 2 c p_c \right].
\label{Ham3}
\end{eqnarray}
With this simplified Hamiltonian and the Poisson bracket structure of the original phase space variables, one can obtain the equations
of motion for these new rescaled variables whose solutions for exponentiated time variable $(t=e^T)$ appear as
\begin{eqnarray}
&& b(t) = \pm \gamma \sqrt{2m/t - 1}, \nonumber \\%\hspace{2cm} 
&& p_b(t) = p_b^{0} \sqrt{t(2 m - t)}, \nonumber \\
 && c(t) = \mp \gamma m p_b^{0} t^{-2}, \nonumber \\%\hspace{3cm} 
 && p_c(t) = \pm t^2,
 \label{Sol.1}
\end{eqnarray}
with $m=t_0$ as an integration constant.
The above solution gives the following line element corresponding  to $q_{ab}$
 \begin{eqnarray}
ds^2=- \frac{dt^2}{\frac{2 m}{t} -1} + \frac{(p_b^{0})^{2}}{L_0^2} \left(\frac{2 m}{t} -1\right) dx^2 + t^2 
d \Omega^2.
\label{line-element2}
\end{eqnarray}
If one chooses the dimension of the integration cell $L_0=p_b^0$, this line element gives the interior of the Schwarzschild spacetime, and also the 
exterior when one moves pass $t=2m$ surface beyond which $t$ appears as a spatial variable.

\subsection{Moving to the quantum corrected  version}
For taking account of the quantum nature one then polymerizes the connections, i.e. uses the holonomies, $h^{(\delta)}[{\cal A}],$ of 
the connection ${\cal A}$ (parameterized by $\delta$) instead of the connection itself and obtain one parameter family of the Hamiltonian constraint.
The parameter $\delta$ captures the notion of the length of the path along which the connection is integrated to obtain the corresponding holonomy.
In this way the Hamiltonian constraint is also obtained up to the polymer parameter, in fact up to a parametric function $\Delta(\delta)$. All the 
elements of this family of parameterized constraints are required to reduce to the classical expression of the Hamiltonian constraint in the
limit $\delta\rightarrow 0$. In terms of the canonical variable one can decide to make use of variable set
$\{ \frac{\sin (\Delta(\delta) \delta b)}{\delta},\frac{\sin \delta c}{\delta}\}$ instead of $\{b,c\}$. Thus, the Hamiltonian constraint 
in the gauge $N = (\gamma \sqrt{|p_c|} \mbox{sgn}(p_c) \delta)/(\sin \Delta(\delta) \delta b)$ becomes
\begin{eqnarray}
\mathcal{C}_{\Delta(\delta)} = - \frac{1}{2 \gamma G_N} 
\Big\{ 2 \frac{\sin \delta c}{\delta}  \  p_c +\left(\frac{\sin \Delta(\delta)\delta b}{\delta} + 
\frac{\gamma^2 \delta }{\sin \Delta(\delta) \delta b} \right) 
p_b \Big\}.
\label{FixN}
\end{eqnarray}
Again, using the Hamiltonian constraint and the equations of motion one obtains
\begin{eqnarray}
 && c(t) =  \frac{2}{\delta} \arctan \Big( \mp \frac{\gamma \delta m p_b^{0}}{2 t^2}  \Big),
\nonumber \\
&& p_c (t) = \pm 
\frac{1}{t^2} 
 \Big[\Big(\frac{\gamma \delta m p_b^{0}}{2}\Big)^2  + t^4 \Big], \nonumber \\
&& \cos \Delta(\delta) \delta b =  \rho(\delta)
\left[ \frac{ 1 - \Big(\frac{ 2 m}{t} \Big)^{\Delta(\delta) \rho(\delta)}
\mathcal{P}(\delta) }
{ 1 + \Big(\frac{ 2 m}{t} \Big)^{\Delta(\delta) \rho(\delta)}
\mathcal{P}(\delta)}
\right], \nonumber \\
&& p_b(t) = -  \frac{2 \ \sin \delta c \
\sin \Delta(\delta) \delta b \ p_c }{\sin^2 \Delta(\delta) \delta b + \gamma^2 \delta^2},
\label{Sol.cpcbpb}
\end{eqnarray}
where,
\begin{eqnarray}
&& \rho(\delta) = \sqrt{1 + \gamma^2 \delta^2}, \nonumber \\
&& \mathcal{P}(\delta) = \frac{\sqrt{1 + \gamma^2 \delta^2} -1}{\sqrt{1 + \gamma^2 \delta^2} +1}.
\label{oro}
\end{eqnarray}
Given these variables one can again construct the spacetime metric which will be bearing some implication of introducing the polymerization
in terms of dependence over $\delta$. Furthermore, as we can see that for $\delta\rightarrow 0$, everything reduces to the previously discussed
classical set-up, thereby assuring of the Schwarzschild limit when the polymer parameter $\delta$ is insignificant.
Therefore, in the case where the quantum correction is expected to be sub-dominant one is hopeful of obtaining the Schwarzschild metric back. 
Moreover, one also requires an asymptotically flat limit of this polymer corrected spacetime. 
One can verify that there exists only one such parametric function $\Delta(\delta)$ which ensures 
the asymptotic limit \cite{modesto1}. This choice for $\Delta(\delta)$ happens to be 
$$\Delta(\delta)=\frac{1}{\sqrt{1+\gamma^2 \delta^2}}.$$
This modified spacetime has some salient features such as existence of two horizons, finiteness of the  Kretschmann scalar with an asymptotic Schwarzschild behavior. For demanding mass independence of the location of Kretschmann scalar peak, one introduces a parameter $a_0$
of the full theory which is the minimum area gap in LQG. For a detailed discussions please refer to \cite{modesto1}.

This approach of polymerizing the connection variable can in principle be implemented in two ways, namely where one polymerizes one of the component or both
$(b,c)$. The former is named as semi-polymeric treatment while the the second one is the full polymeric treatment.

We consider the full polymeric loop black hole in this paper. This singularity free semiclassical black hole metric is given by
\begin{equation}
 ds^2=-f(r)dt^2+g(r) dr^2+h(r)r^2 d\Omega^2, \label{FPMet}
\end{equation}
where,
$$f(r)=\frac{(r-r_+)(r-r_-)(r+r_*)^2}{r^4+a_0^2},$$
$$g(r)=\left\lbrace \frac{(r-r_+)(r-r_-)r^4}{(r+r_*)^2(r^4+a_0^2)}\right\rbrace ^{-1},$$
$$h(r)=1+\frac{a_0^2}{r^4},$$
where $a_0$ is the lowest area gap quanta in LQG,
$r_+=2m$, $r_-=2m{\cal P}^{2}$ and $r_*=\sqrt{r_{+}r_{-}}=2m{\cal P}$,
with the quantity $${\cal P}=\frac{\sqrt{\epsilon^2+1}-1}{\sqrt{\epsilon^2+1}+1},$$ being the polymeric function 
which captures the correction from
LQG in terms of $\epsilon=\gamma \delta$.
Qualitatively, this spacetime has similar horizon structure as that of a  Reissner-Nordstrom black holes having two horizons at $r_-$ and $r_+$. Another important aspect of this black hole solution is that it is self-dual in sense of T-duality. One can verify that under the transformation
$r\rightarrow a_0/r$, the metric remains invariant, with suitable parameterization of other variables (details in  \cite{Modesto:2009ve}), hence marking itself as its dual description under T-duality .

As we have argued before, the correction to observables because of $\epsilon$ is of our interest and the correction due to $a_0$ is safely neglected; $a_0$ being a dimensionful quantity and hence appearing in observables with high energy scale suppression. So we will henceforth put $a_0$ to zero. Some interesting properties and phenomenology of this Loop Quantum Black Hole (LQBH) is explored in  \cite{Modesto:2009ve}.
\section{Weak Field Lensing} \label{weak} 
First let us consider lensing in the small deflection limit.
If we expand the metric in the weak field limit, we get 
\be
f(r)=1-\frac{2m}{r}(1-{\cal P})^{2}+\frac{8m^{2}}{r^{2}}(-{\cal P}+{\cal P}^{2}-{\cal P}^{3}),
\ee
 and
 \be
g(r)=1+\frac{2m}{r}(1+{\cal P})^{2}+\frac{4m^{2}}{r^{2}}(1+2{\cal P}+3{\cal P}^{2}+2{\cal P}^{3}),
\ee
 from which we can read off the ADM mass of the system $M$ as 
 \be M=m(1+{\cal P})^2. \ee
 
 Rewriting the weak field expansion in terms of  Arnowitt-Deser-Misner (ADM) mass and using the formalism of Keeton and Peeters \cite{Keeton:2005jd} to identify  the parametrized post-Newtonian (PPN) parameters, we arrive at the expansion form of the deflection angle in terms of the impact parameter $b$ 
 \be \hat{\alpha}(b)=A_{1}\left(\frac{M}{b}\right)+A_{2}\left(\frac{M}{b}\rt)^{2}+...,
 \ee
 where the parameters in the expansion are given as
  \be A_{1}=4-8{\cal P}\mbox{ and \ensuremath{A_{2}=\pi(\frac{15}{4}-18{\cal P})}}, \ee
  keeping terms only up to first order in ${\cal P}$ since the polymeric function, 
$${\cal P}=\frac{\sqrt{1+\epsilon^2} -1}{\sqrt{1+\epsilon^2} +1}\sim{\cal O}(\epsilon^2)\ll 1.$$
  
 Precision solar system observations, more specifically solar gravitational
deflection of Radio Waves using Geodetic Very-Long-Baseline Interferometry
Data \cite{PhysRevLett.92.121101} put very strong constraints on $A_{1}$ (the constraints on $A_2$ etc is very loose) as $A_{1}=3.99966\pm0.00090$. Thus $A_1$ is constrained to be extremely close to 4. This constraint can be expressed in terms of $\epsilon$ (which we recall is the product of $\delta$ and Immirzi parameter $\gamma$) and since the Immirzi parameter is estimated to be $\gamma\sim0.25$ by demanding that area law holds in loop quantum gravity (for example see \cite{Lochan:2012in,Rovelli:1996dv}) one gets
\be
\delta\lesssim0.1.
\label{delcon} 
\ee

It is surprising that solar system experiments have anything useful at all to say about  a quantum
gravity motivated parameter. This is because of a
certain peculiarity of the model under consideration alluded to before (viz, the dimensionlessness
of a quantum gravity motivated correction) which we discuss further
later in this paper. We can also estimate the correction to various
weak field observables (say the far field Einstein Ring (ER) location). Let $\theta_{E_{0}}$ denote
the angular location of ER for Schwarzschild metric. Then $$\theta_{E_{0}}=\sqrt{\frac{4Md_{ls}}{d_{ol}d_{os}}},$$ where $d_{ls}$ is distance between lens and source, $d_{os}$ is distance between observer and source and $d_{ol}$ is distance between lens and observer.
Using zeroth order solution to the lens equation following \cite{Keeton:2005jd} we
get the angular ER location for loop correction as 
\begin{equation}
\theta_{E}=\theta_{E_{0}}\sqrt{\frac{A_1}{4}}\simeq\theta_{E_{0}}(1-{\cal P}),\label{fe} 
\end{equation}
 up to first order in ${\cal P}$. Thus ER shrinks slightly as compared to the Schwarzschild solution.

There is a caveat however. We have assumed that the metric which was
derived as a solution for black hole spacetime is a valid description
for the vacuum spacetime geometry far away from a compact (horizonless)
object in the theory. This is a non-trivial assumption but a plausible
one. If these black holes can be formed via spherically symmetric
gravitational collapse in the theory, starting from a compact object,
the spacetime outside the object should not change in the process
as the information about the collapse can not be communicated to asymptotia
because of impossibility of monopolar gravitational wave generation.
And so if the consequent black hole spacetime is unique the outside
vacuum of a compact object must be described by the same spacetime
asymptotically as that describing the black hole asymptotia. So with this working proposition, the equation \ref{delcon} gives us a constraint on polymeric parameter $\delta$ from solar system tests.

\section{Strong field lensing}\label{strong} 
Now we turn to lensing in strong field lensing and explore the properties of relativistic images.
\subsection{The formalism: strong lensing coefficients and observables}
It has been known since a long time that deflection angle of photons can become arbitrarily large (in principle infinite), if the photons are allowed to explore regions close enough to the black hole \cite{dar59,dar61}. So photons can loop around the black hole multiple times on the way from source to observer leading to the  phenomenon of relativistic lensing and formation of relativistic images \cite{Virbhadra:2000ju,nsl.ve}. The observation that the deflection angle diverges logarithmically as the distance of closest approach approaches the radius of the photon sphere has been used to define and compute the so called strong lensing coefficients and strong lensing observables by Bozza \cite{Bozza:2002zj}. We use this formalism for our purposes.
% and briefly review it below.
Since higher order images are clumped together, Bozza \cite{Bozza:2002zj} considered a situation where one is able to resolve only the first relativistic image from the rest 
and compare the brightness of the first image with the cumulative brightness of all other relativistic images. The relavant strong lensing observables are then $s$, the separation between the outermost image and rest of the images; $r$, the luminosity ratio of outermost image and rest of the images (clumped together); and $\theta_{\infty}$, the angular location of black hole shadow. Details of the calculation of strong lensing coefficients and observables is given in Appendix A.

\subsection{Strong lensing coefficients and observables for LQG Black Hole }
\begin{figure}[htp]
  \includegraphics[width=6cm]{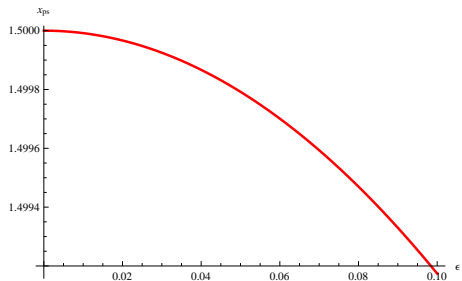}
 \caption{The radius of photon sphere (scaled by Schwarzschild radius) as a function of $\epsilon$ characterizing quantum gravity corrections} 
 \label{fig1}
\end{figure}

As we have seen earlier in \ref{FPMet}, 
in the Schwarzschild units the spacetime has double horizon at $r_+=1$ and $r_-={\cal P}^2$. This spacetime contains two
photon spheres. The outer one is the unstable photon sphere while the inner horizon
is the stable one. The radius of outer photon sphere (which is the one relevant for lensing) decreases with the increasing $\epsilon$ as shown in \ref{fig1}. 

\begin{figure}[ht]

\begin{center}
 \includegraphics[width=6cm]{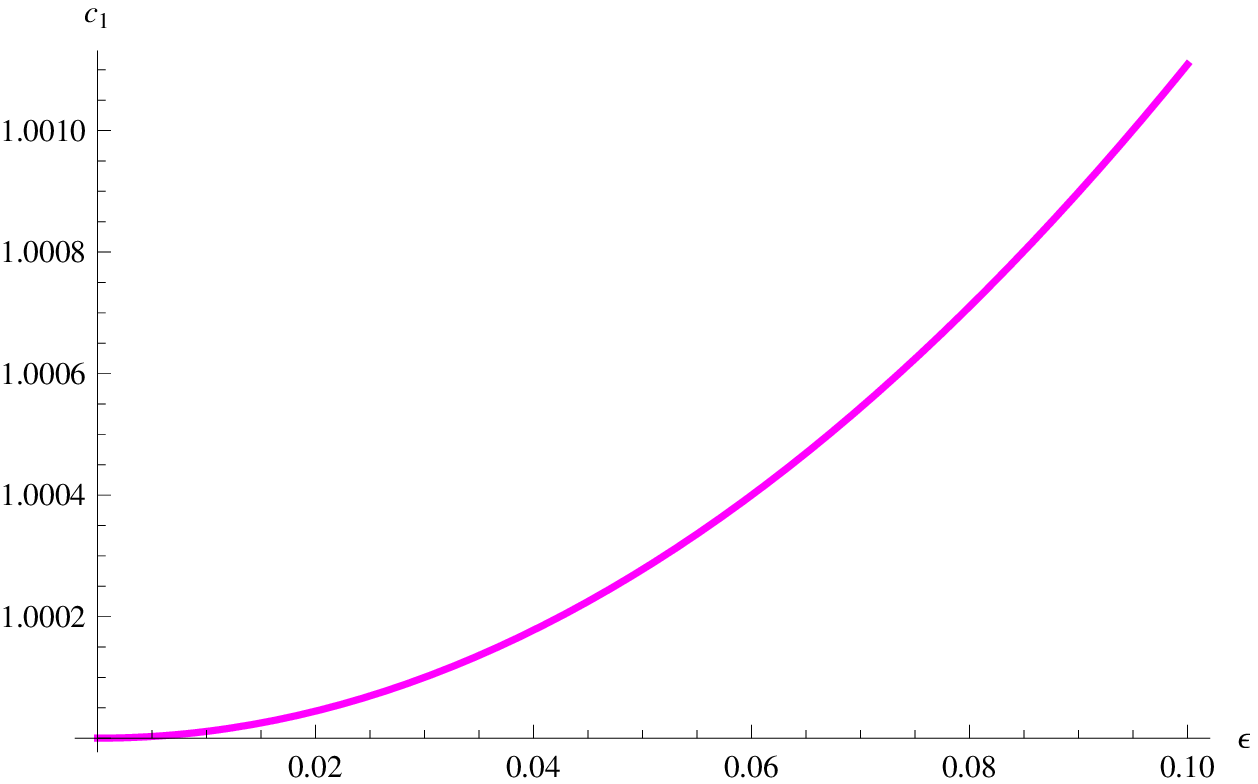}
 % c1lqg.eps: 0x0 pixel, 300dpi, 0.00x0.00 cm, bb=
\includegraphics[width=6cm]{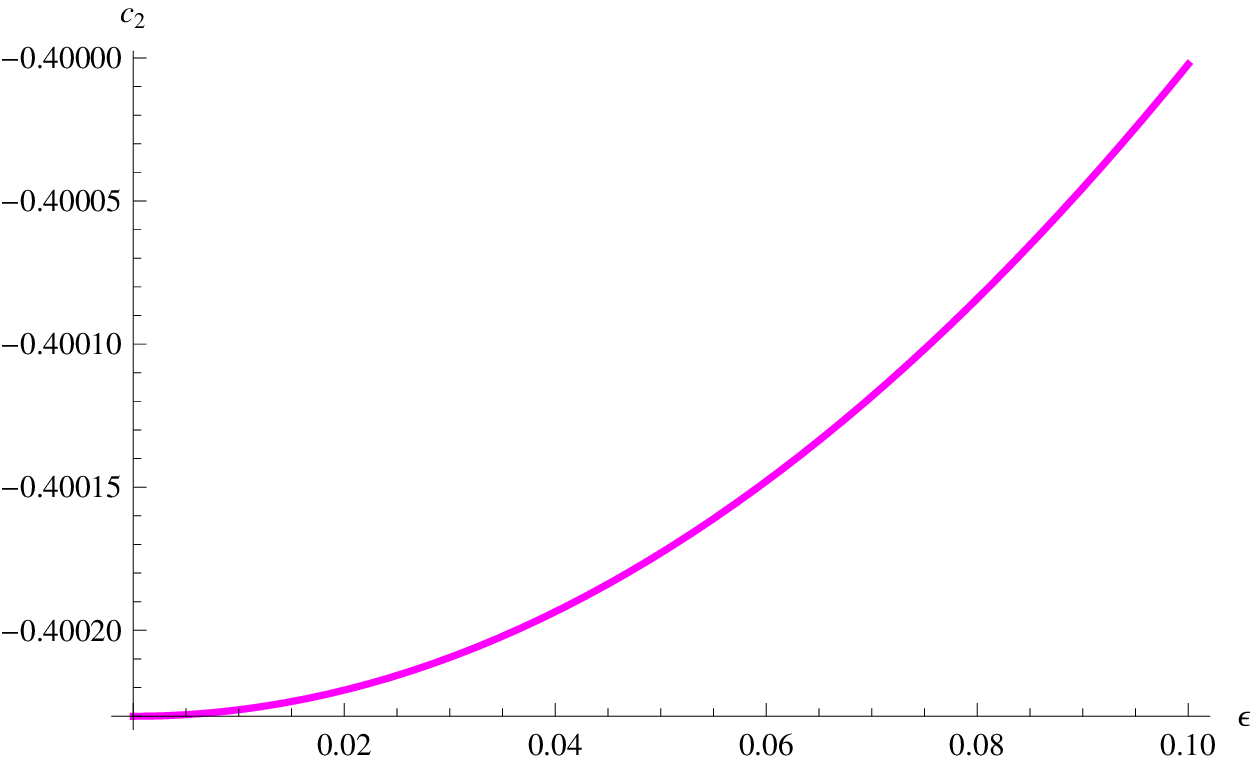}\\
% c2plqg.eps: 0x0 pixel, 300dpi, 0.00x0.00 cm, bb=
\includegraphics[width=6cm]{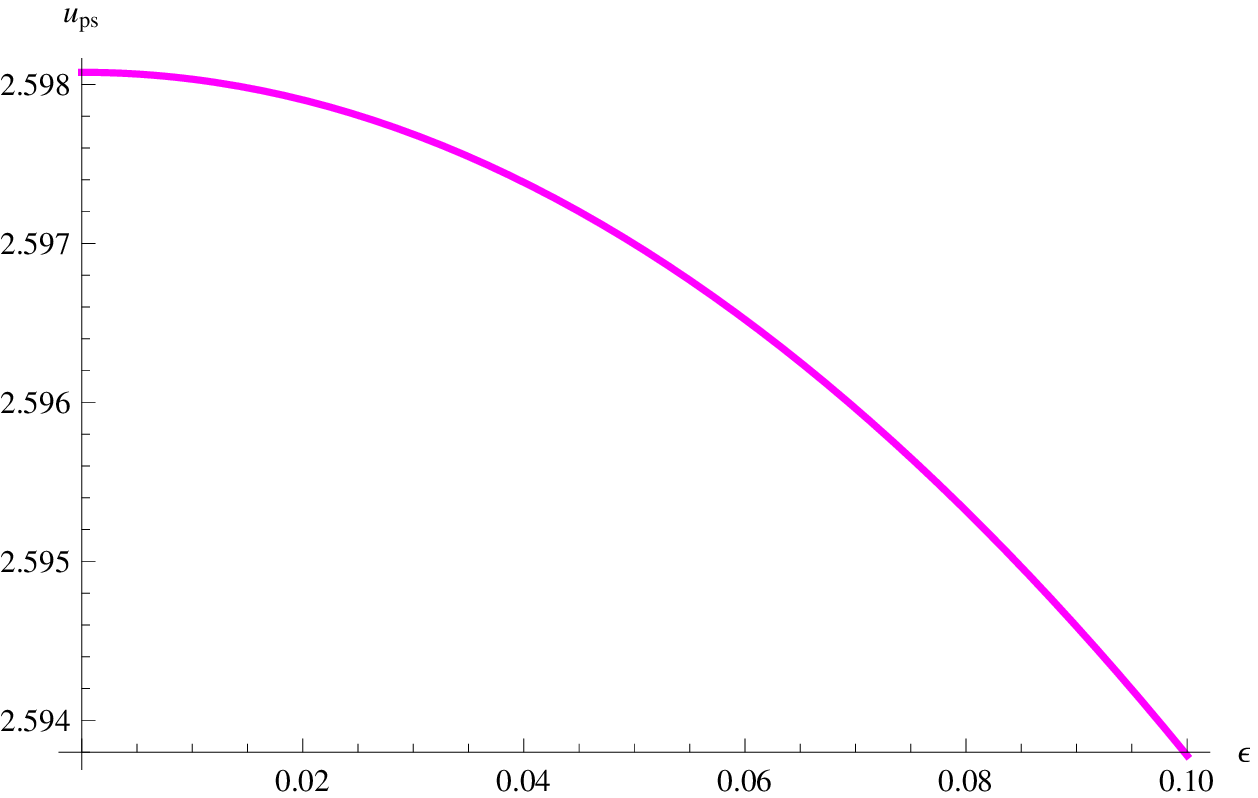}
% umplqg.eps: 0x0 pixel, 300dpi, 0.00x0.00 cm, bb=
\end{center}

 \caption{The strong lensing coefficients as a function of $\epsilon$ characterizing quantum gravity corrections} 
 \label{fig2}
\end{figure}

As the radius of photon sphere decreases with the loop correction so does the corresponding impact parameter. Computation of  the strong lensing coefficients shows that that $c_1$ and $c_2$  increase monotonically while $u_{ps}$ decreases monotonically  as the quantum correction
grows stronger (\ref{fig2}). 

At linear order in ${\cal P}$, corrections to strong lensing coefficients  can be calculated as

\begin{eqnarray}
c_{1}&=&1+\frac{4{\cal P}}{9} ,\\
c_{2}&=&(1+\frac{4{\cal P}}{9})\ln\left(\frac{3}{2}+\frac{4{\cal P}}{3}\right)+(-0.40+0.09{\cal P}),\\
u_{ps}&=&\frac{3\sqrt{3}}{2}\left( 1-\frac{76{\cal P}}{9}\right),  \label{nefa}\\
a_{R}&=&2\ln(6(2-\sqrt{3}))+\frac{8\sqrt{3}{\cal P}}{9(3+\sqrt{3})}\ln(6(2-\sqrt{3}))-\frac{8{\cal P}[9+\sqrt{3}+\ln(\frac{26+15\sqrt{3}}{216})]}{9(3+\sqrt{3})}.
\end{eqnarray}

As expected ${\cal P}\rightarrow0$ reproduces the Schwarzschild values.

\begin{figure}[ht]
\begin{center}
 \includegraphics[width=6cm]{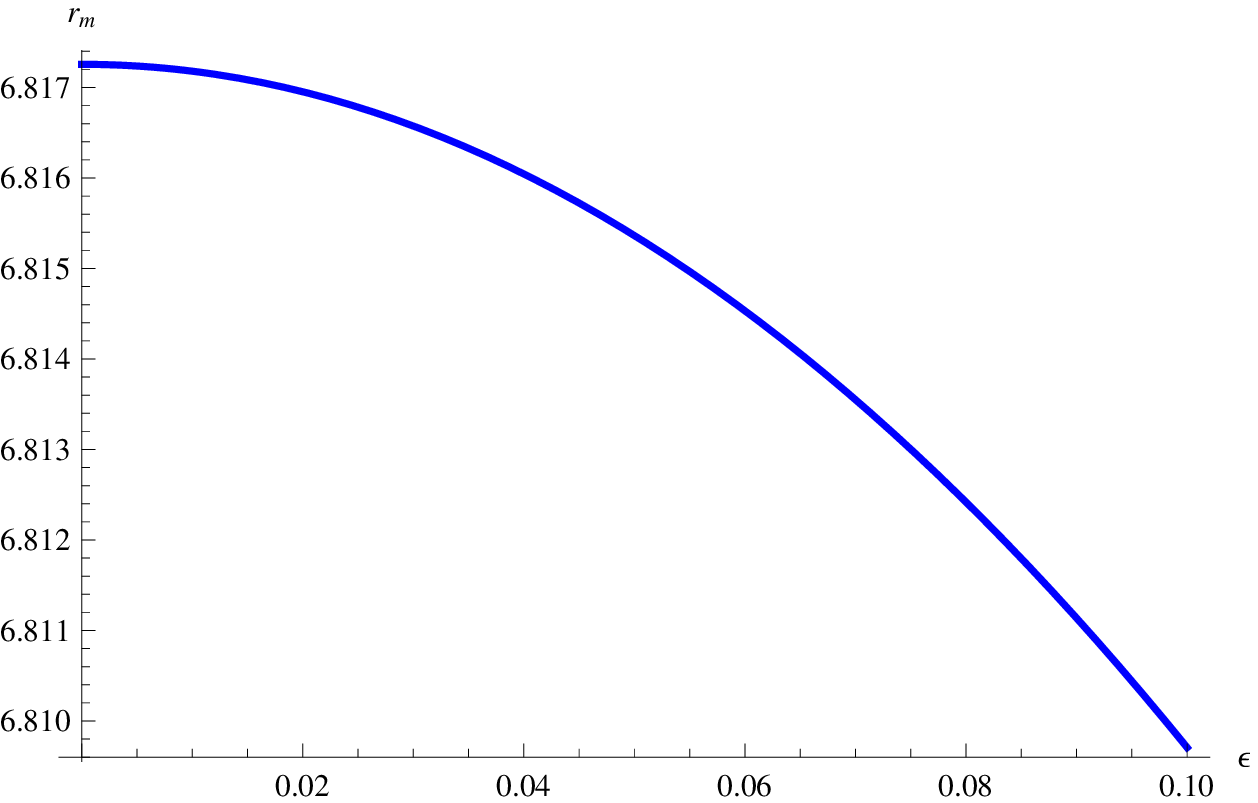}
 % c1lqg.eps: 0x0 pixel, 300dpi, 0.00x0.00 cm, bb=
\includegraphics[width=6cm]{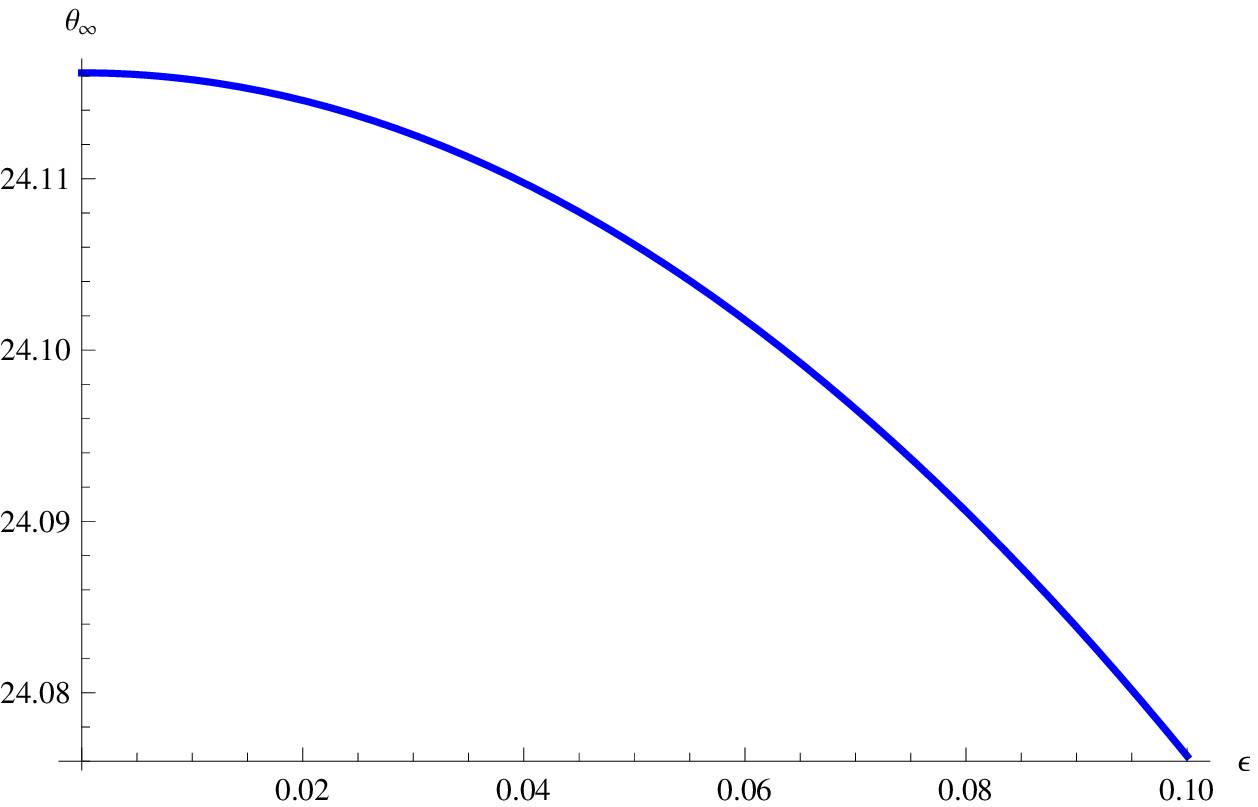}\\
% c2plqg.eps: 0x0 pixel, 300dpi, 0.00x0.00 cm, bb=
\includegraphics[width=6cm]{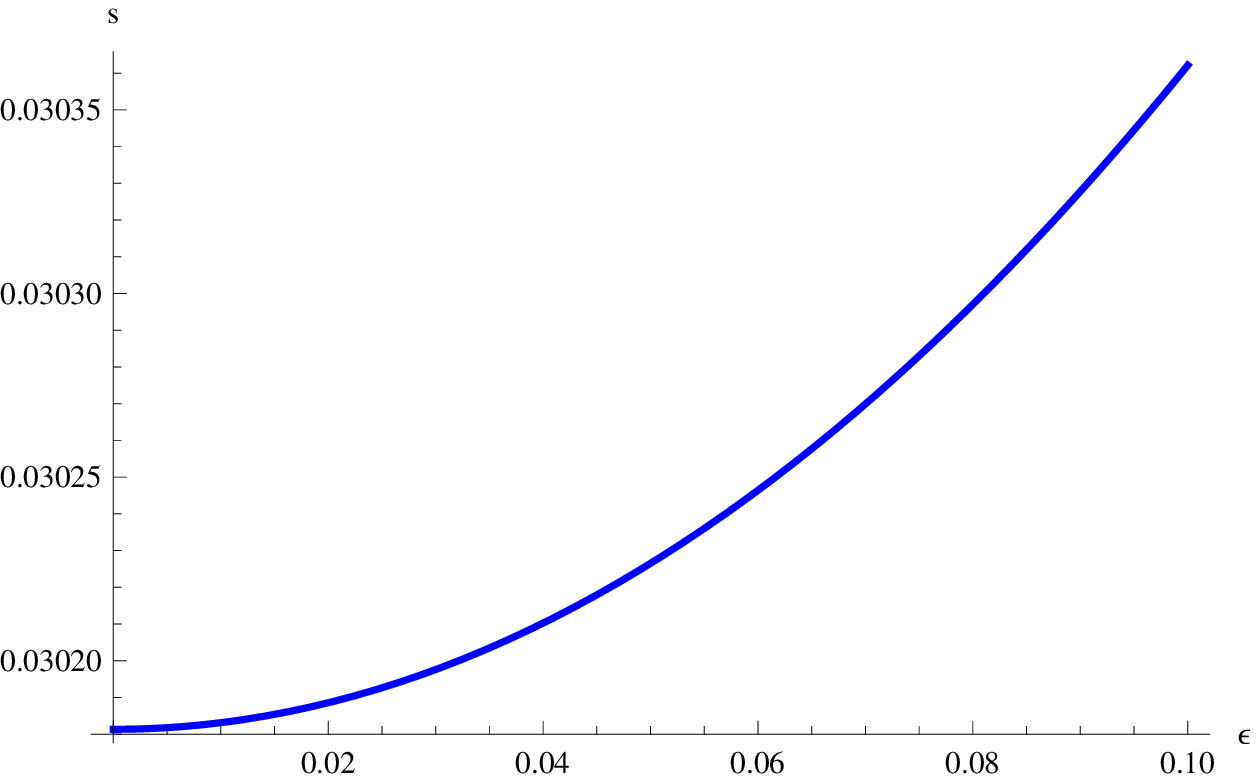}
% umplqg.eps: 0x0 pixel, 300dpi, 0.00x0.00 cm, bb=

\end{center}
 \caption{The strong lensing observables as a function of $\epsilon$ characterizing quantum gravity correction for the galactic super-massive black hole scenario. $r$ is in magnitude and $\theta_{\infty}$ and $s$ are in microarcseconds. } 
 \label{fig3}
\end{figure}

\begin{figure}[htp]
 \centering
  \includegraphics[width=6cm]{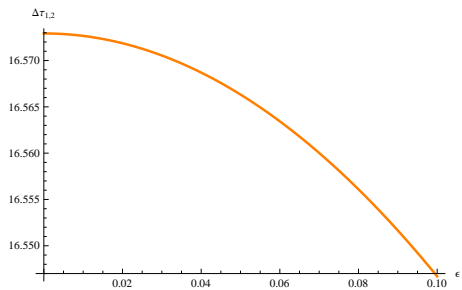}
  \caption{The time delay (in dimensionless units) between first and second relativistic Einstein rings  a function of $\epsilon$ characterizing quantum gravity correction}
 \label{fig4}

\end{figure}

We now use these coefficients to compute the observational quantities  $r$, $\theta_{\infty}$ and $s$ and plot their behavior with the increasing
strength of the polymeric parameter in \ref{fig3}. 
The strong lensing observables have been computed for the galactic super-massive black hole scenario 
($   M\simeq4\times10^{6}M_{\odot}, d_{ol}\simeq8.5kpc                                  $ and with  $d_{os}=2d_{ol}$  ) as a concrete case.
So, we see that with increasing polymeric parameter, the angular location of shadow region decreases slightly, the separation between the first and the last images increases, while the luminosity ratio decreases. The fractional change is surely insignificant for practical purposes as for as one can foresee the limits of current and proposed observational technologies. However situation for time delay observations is more promising, as we shall discuss in next section. The time delay between successive relativistic Einstein rings is given by \cite{Bozza:2003cp} (in Schwarzschild units, i.e., scaled by $2M$) 
\begin{equation}
\Delta\tau_{N,N+1}=2\pi u_{ph}+2\frac{\omega_{ps}}{\chi_{ps}}\sqrt{\frac{u_{ps}}{c_{ps}}}e^{\frac{c_{2}}{2c_{1}}}e^{\frac{-N\pi}{c_{1}}}\left(1-e^{\frac{-\pi}{c_{1}}}\right),
\end{equation}
where
\begin{equation}
c=\frac{2\chi-\chi''x^{2}}{4\chi}\frac{1}{x\sqrt{\chi}}.
\end{equation}
To zeroth order time delay between successive relativistic images is simply $2\pi u_{ps} \times 2M$ and so as the critical impact parameter $u_{ps}$ decreases so does $\Delta\tau_{1,2}$ as shown in \ref{fig4}.

\section{Observational prospects}\label{obs} 

Since the factional change in angular location of photon sphere, which can be read off from \ref{nefa} is $\sim\frac{76{\cal P}}{9}$, the resolution needed for observing such a tiny change in relativistic images for $\delta\sim0.1$ will be tens of nanoarcseconds as the typical angular location of the relativistic 
images in our galaxy is tens of microarcseconds ($\sim24\mu''$). This looks more optimistic
if we consider billion solar mass black holes in other galactic centers
as the radius of photon sphere will accordingly scale up. This will of course be compensated by distance
factor to other galaxies when we consider the angular resolutions
needed. However the time delay between relativistic images will be immune to
the distance factor since time
delays between successive images is roughly the is roughly the time taken by light to
travel one closed orbit on the photon sphere. In fact it is well approximated by $2\pi u_{ps}$ \cite{Bozza:2003cp}. So we can write
\be 
\overline{\Delta T}=\Delta T(1-\frac{76{\cal P}}{9})
\ee
using $u_{ps}$ from \ref{nefa} 
where $\Delta T$ is time delay between relativistic images for Schwarzschild black hole and $\overline{\Delta T}$
is analogous quantity for loop black hole. The required time delay resolution to probe polymeric corrections for certain billion solar mass
black holes can be of the order of seconds (assuming $\delta\sim0.1$) as the time delay for them would be in hours \cite{Bozza:2003cp}.

\section{Contrast with other quantum gravity scenarios}
%:\textit{the dimensionlessness of ${\cal P}$} }

\label{cont} 

A notable feature of the quantum gravity motivated metric we study is
that one of the parameters that characterize the deviation from Schwarzschild geometry is dimensionless.
This is peculiar and is in contrast to other quantum gravity motivated scenarios
like non-commutative black hole space time \cite{Nicolini:2005vd} where the non-commutativity parameter has dimension
of $M^2$, or like the Callan-Myers-Perry black hole (CMP) \cite{Callan:1988hs} in String theory where the string tension
has dimension of $M^2$ as well. In those scenarios non-commutative parameter and string
tension set up a scale and all corrections are mass suppressed. In other words, if the quantum gravity motivated dimensionful parameter is $l_{qg}$  then the corrections
to observables generically  takes the form $\left(\frac{l_{qg}}{M}\right)^\mathscr{m}$  where $\mathscr{m}$ is some positive power and $M$ is the mass of the black hole. This mass suppression
effect is absent in the space time under study and makes it more amenable
to observational tests compared to other generic quantum gravity corrections such
as the ones discussed before. Ignoring this mass suppression, however,
strong lensing in non-commutative black hole space times \cite{Ding:2010dc} shows similar trend as the space time under consideration, and it might be interesting to investigate
the effect of string tension corrections in CMP solution for strong lensing which is beyond the scope of current study. 

\section{Einstein Ring system as a probe: Consistency relations}\label{erc} 
In a given model for space time describing a black hole one can look
for useful relations between strong field and weak field observations
and one can construct certain consistency relations. For example one
can relate the far field ER with the relativistic ERs (which are clumped near the angular location of photon sphere \cite{Bozza:2002zj}). This was used in \cite{Tsukamoto:2012xs}  to distinguish black hole and wormhole systems. This idea can be used to obtain consistency
relation for strong and weak field lensing in observables for our model as below.

Let $\theta_{0}$ and $\theta_{\infty}$ be the far field and relativistic
ERs for Schwarzschild spacetime and $\bar{\theta}_{0}$ and $\bar{\theta}_{\infty}$
be corresponding quantities for the loop black hole. Then one can show that \cite{Tsukamoto:2012xs}
\be
\theta_{\infty}=\frac{3\sqrt{3}}{4}\frac{d_{os}}{d_{ls}}\theta_{0}^{2}.
\ee
Since
we have
\be \overline{\theta}_{0}=\theta_{0}(1-{\cal P}),\ee
from \ref{fe} and 
\be \overline{\theta}_{\infty}=\theta_{\infty}(1-\frac{76{\cal P}}{9}),\ee using the expression for $u_{ps}$ from \ref{nefa} 
the consistency
relation becomes \be \overline{\theta}_{\infty}=\frac{3\sqrt{3}}{4}\frac{d_{os}}{d_{ls}}\overline{\theta}_{0}^{2}\left(1-\frac{58{\cal P}}{9}\right),\ee
for loop black hole. A deviation from this relation will invalidate the model. Since ${\cal P}$
is expected to be an universal constant, this gives a line in $(\overline{\theta}_{0},\overline{\theta}_{\infty})$ plane where all observations must lie to be consistent with the model.

\section{Discussion and Conclusion}\label{diss} 

The gravitational lensing signatures in both strong and weak deflection regimes were studied for some black hole solutions in LQG recently obtained by Modesto \cite{modesto1}. We
highlighted a curious feature of the solution, viz., the appearance of a dimensionless parameter (related to Barbero–Immirzi parameter and polymeric parameter) in the solution under study. This is in contrast to generic quantum gravity corrections such as the noncommutativity parameter or the string tension or the minimal area quanta etc. which are always dimensionful. This helps the solution to evade the mass suppression effects, i.e., the observables are corrected not by powers of ratio of quantum gravity scale to mass of black hole but by powers of the dimensionless parameter itself. By appealing to some general principles, we used the far away regions for this spacetime to model the spacetime away from  any spherically symmetric matter distribution in the theory far away from central source. In particular then, this can be used to calculate corrections to observables in solar system observations. When used in conjunction with theoretically motivated values  of Barbero-Immirzi parameter one can use this to constrain polymeric parameter from precision solar system observations. We find that the polymeric parameter $\delta$ has to be less than $\sim0.1$.
We also computed the strong lensing observables and found  that with increasing polymeric parameter, the angular location of shadow region decreases slightly, the separation between the first and the last images increases while the luminosity ratio decreases. Strong lensing coefficients are interesting even in themselves because of their relationship with
quasinormal modes \cite{Stefanov:2010xz} and absorption cross sections from black holes \cite{Wei:2011zw}. Quasinormal modes have, for example, interesting link to semi-classical aspects of black hole physics \cite{Maggiore:2007nq} and for quantum gravity motivated solutions like the one studied in this paper. Therefore strong lensing coefficients might provide interesting theoretical motivations for further exploration. 

From observational vantage point, the magnitude of these effects is too small to be detected or ruled out  in near future observations for galactic super-massive black holes. However, the required time delay resolution to probe polymeric corrections for certain billion solar mass black holes in distant galaxies can be of the order of seconds and seems a promising probe of these corrections once relativistic lensing observations become a reality. 
Even if we do not have the requisite angular resolution to distinguish between Schwarzschild and Loop Quantum Gravity motivated black holes, time delay observations might still be able to discriminate between these models. Important point to note is that this kind of quantum gravity correction has a reasonable chance of being probed in future astrophysical observations while noncommutativity parameter or string tension etc. have no realistic chance whatsoever because of the mass suppression effect. One caveat however is that any deviation from the vacuum solution in general relativity (for example black holes in Brans-Dicke theory etc.) would have qualitatively similar deviations from Schwarzschild strong and weak field lensing observables (although the exact strong-weak consistency relations would  be quantitatively different) and one needs to be careful in interpreting the deviation from Schwarzschild spacetime. But the models having the feature that the leading power of the parameters, in the expressions capturing the deviation (from their corresponding Schwarzschild values), is different in strong and weak field regimes (unlike Brans-Dicke where the correction shows up at linear order for both strong and weak field regimes) can be more easily distinguished from this LQG motivated model.

\begin{appendices}
\section{: Bozza's formalism for computing strong lensing observables}

The  general spherically symmetric metric can be written in the form given by equation \ref{FPMet} where $f(r)$, $g(r)$ and $h(r)$ are arbitrary functions. We also restrict them to be such that the spacetime is asymptotically flat.
% \begin{equation}
%  ds^{2}=-f(r)dt^{2}+g(r)dr^{2}+h(r)r^2 d\Omega^{2}.
% \end{equation}

For convenience we define two new functions in terms of metric coefficients 
$$\omega(r)=\frac{\sqrt{f(r)g(r)}}{h(r)},$$
$$\chi(r)=\frac{f(r)}{h(r)}.$$ 
The photon sphere which describes a null geodesic at a fixed co-ordinate radius $r$ is obtained by solving the 
equation 
\begin{equation}
r\chi'(r)=2 \chi
\end{equation}
in terms of one of these new functions. The deflection angle which gives total angular deflection of a null geodesic
coming from and going to infinity is given as
\begin{equation}
 \alpha (r_{0})=I(r_{0})-\pi ,
\label{sl4}
\end{equation}
where $r_0$ is the closest approach distance and
\begin{equation}
I(r_{0})=2\int_{r_{0}}^{\infty }\left[ \frac
{g(r)}{h(r)r^2}\right] ^{1/2} \left[ \frac
{h(r)r^2f(r_{0})}{h(r_{0})r_0^2f(r)}-1\right] ^{-1/2}dr. \label{IR}
\end{equation}
Let us define a dimensionless variable $z=\frac{r_0}{r}$ and $\eta=1-z$. In terms of these new variables
and the new functions described above, the expression \ref{IR} gets simplified to
\begin{equation}
I(r_{0})=2\int_{0}^{1}\frac{\omega(r_{0}/z)} {\sqrt{\chi
(r_{0})-\chi
(r_{0}/z)z^{2}}}dz=2\int_{0}^{1}\frac{\omega(r_{0}/(1-\eta))}
{\sqrt{\chi (r_{0})-\chi (r_{0}/(1-\eta))(1-\eta)^{2}}}d\eta.
\end{equation}
We will be interested in the divergent structure of the integrand around $r=r_0$ which will give us a measure of the 
strong lensing coefficients, as we will see. Expanding the integrand in leading orders of $\eta$ we separate out the
integral into a divergent and a regular part as
\begin{equation}
I(r_{0})=I_{D}(r_{0})+I_{R}(r_{0}), \label{sl9}
\end{equation}
with
\begin{equation}
I_{D}(r_{0})=2\int_{0}^{1}\frac{\omega(r_0)}
{\sqrt{\gamma_{1}(r_{0})(1-z)+\gamma_{2}(r_{0})(1-z)^{2}}}dz,
\label{sl10}
\end{equation} 
and
\begin{equation} I_{R}(r_{0})=2\int_{0}^{1}\left[ \frac{\omega(r_{0}/z)} {\sqrt{\chi
(r_{0})-\chi(r_{0}/z)z^{2}}}-\frac{\omega(r_0)}
{\sqrt{\gamma_{1}(r_{0})(1-z)+\gamma_{2}(r_{0})(1-z)^{2}}}\right]
dz. \label{sl11}
\end{equation}
In the above expressions
$$\gamma_{1}(r_{0})=2\chi (r_{0})-r_{0}\chi '(r_{0}),$$ and
$$\gamma_{2}(r_{0})=-\chi (r_{0})+r_{0}\chi '(r_{0})-(1/2)r_{0}^{2}
\chi ''(r_{0}).$$
 The integral $I_{D}(r_{0})$ can be calculated exactly to
give
\begin{equation}
I_{D}(r_{0})=-\frac{2 \omega(r_0) } {\sqrt{\gamma_{2}(r_{0})}} \ln
\frac{\gamma_{1}(r_{0})} {\left[
\sqrt{\gamma_{1}(r_{0})+\gamma_{2}(r_{0})}+\sqrt{\gamma_{2}(r_{0})}
\right]^{2}}, \label{sl12} 
\end{equation}
which diverges for the  $r_{0}=r_{ps}$, with $r_{ps}$ being the radius of photon sphere. For $r_0$ close to $r_{ps}$
we can again expand $I_{D}(r_{0})$, we can write
\begin{equation}
I_{D}(r_{0})=-\frac{\sqrt{8} \omega(r_{ps})}{ \sqrt{2 \chi
(r_{ps})-r_{ps}^{2}\chi ''(r_{ps})}} \left[ \ln\left(
\frac{r_{0}}{r_{ps}}-1 \right) -\ln 2 \right] +O(r_{0}-r_{ps}),
\label{sl13}
\end{equation}
 using  $2\chi (r_{ps})-r \chi '(r_{ps})=0$.
We also see that $I_{R}(r_{0})$ which converges for $r_{0}=r_{ps}$, can be written as
$$I_{R}(r_{0})=I_{R}(r_{ps})+O(r_{0}-r_{ps}).$$

The deflection angle \ref{sl4}, which diverges for $r_0=r_{ps}$, is
thus given in the strong deflection limit by
\begin{equation}
\alpha(r_{0})=-a_1\ln\left(\frac{r_{0}}{r_{ps}}-1\right)+a_2+O(r_{0}-r_{ps}),
\label{sl14}
\end{equation} where

$$ a_{1}=\frac{\sqrt{8} \omega(r_{ph})} {\sqrt{2\chi
(r_{ps})-r_{ps}^{2}\chi''(r_{ps})}}, \label{sl15}
$$
 $$ a_{2}=-\pi +a_{D}+a_{R}, \label{sl16}
$$
$$ a_{D}=a_{1}\ln{2}, \label{sl17}$$
and
 $$ a_{R}=I_{R}(r_{ps})=2\int_{0}^{1}\left[\frac{\omega(r_{ps}/z)} {\sqrt{\chi
(r_{ps})-\chi (r_{ps}/z)z^{2}}}- \frac{\sqrt{2} \omega(r_{ps}) }
{(1-z)\sqrt{2\chi(r_{ps})-r_{ps}^{2}\chi ''(r_{ps})}}\right]dz.
\label{sl18}
$$
The deflection angle can be obtained as a function of the impact
parameter $u(r_0)=\frac{r_0}{\sqrt{\chi(r_0)}}$ with
\begin{equation}
\alpha (u)=-c_{1}\ln \left( \frac{u}{u_{ps}}-1 \right)
+c_{2}+O(u-u_{ps}), 
\end{equation}
 where $c_{1}=a_{1}/2$ and
\begin{equation}
 c_{2}=\frac{a_{1}}{2}\ln{\frac{2\chi (r_{ps})-r_{ps}^{2}\chi
''(r_{ps})} {4\chi(r_{ps})}}+a_{2}. 
\end{equation}
Further, we can express the deflection angle as a function of image position $\theta$
using $u=d_{ol} \theta$ as
 \begin{equation}
\alpha (\theta)=-c_{1}\ln
\left( \frac{d_{ol}\theta}{u_{ps}}-1 \right) +c_{2}+O(u-u_{ps}),
\label{sl22}
\end{equation}
where $u_{ps}=\frac{r_{ps}}{\sqrt{\chi(r_{ps})}}$ and $d_{ol}$ gives the distance between the observer and the lens.
In the above equation \ref{sl22} the coefficients $c_1$ and $c_2$ are known as the strong lensing coefficients.
They can be related to the lensing observables through the Virbhadra-Ellis lens equation \cite{Virbhadra:2000ju}
\begin{equation} \tan \beta =\tan \theta -c_{3}\left[ \tan
(\alpha -\theta) +\tan \theta \right] , \label{pm1} 
\end{equation}
where $c_{3}=d_{ls}/d_{os}$. In the above expression $\beta$ gives the source position, while $d_{ls}$ and $d_{os}$ are the lens to source and observer to source
distances, respectively. For an aligned configuration we consider $\beta$ and $\theta$ to be small. Whereas, for strong lensing scenario the deflection angle $\alpha$ is
typically of the order $2n\pi$. Thus, $n=0$ corresponds to the weak field limit which gives two weak field limit images while for $n>0$ we get two infinite set of
relativistic images.\\
We substitute 
$$\alpha =2n\pi +\Delta \alpha _{n},$$ with
$n\in \mathbb{N}$ and $0<\Delta \alpha _{n}\ll 1$, for one set of  relativistic images (belonging to the same side of the source), the lens equation gets simplified to 
\begin{equation}
\beta =\theta -c_{3}\Delta \alpha _{n}. \label{pm2}
\end{equation}
The other set of images (on the other side of source) will then be given by
$$\alpha =-2n\pi -\Delta \alpha _{n}.$$
Using \ref{sl22} to the leading order we can invert $\alpha$ in favor of $\theta$ as
\begin{equation}
\theta (\alpha )=\frac{u_{ps}}{d_{ol}}\left[
1+e^{(c_{2}-\alpha )/c_{1}} \right] , \label{pm5}
\end{equation}
We now make a first order Taylor expansion around $\alpha =2n\pi
$, to express the angular position of the $n$-th image ss
\begin{equation}
\theta _{n}=\theta ^{0}_{n}-\zeta _{n}\Delta \alpha _{n},
\label{pm6} 
\end{equation}
 with
\begin{equation}
\theta ^{0}_{n}=\frac{u_{ps}}{d_{ol}}\left[ 1+e^{(c_{2}-2n\pi
)/c_{1}}
 \right] ,
\label{pm7} 
\end{equation}
 and
\begin{equation}
\zeta _{n}=\frac{u_{ps}}{c_{1}d_{ol}}e^{(c_{2}-2n\pi )/c_{1}}.
\label{pm8}
\end{equation}
Solving the Lens equation upto first order gives,
\begin{equation}
 \theta _{n}=\theta
^{0}_{n}+\frac {\zeta _{n}}{c_{3}}(\beta -\theta ^{0}_{n}).
\label{pm14}
\end{equation}
For small $c_1$ and $c_2$ \ref{pm8} suggests that $\zeta_n$ is vanishingly small leading to the fact that 
 all relativistic images lie very close to $\theta ^{0}_{n}$.
Similarly  the other set of relativistic images have
angular positions
\begin{equation}
\theta _{n}=-\theta ^{0}_{n}+\frac {\zeta _{n}}{c_{3}}(\beta
+\theta ^{0}_{n}). \label{pm15} 
\end{equation}
 In the case of perfect alignment ($\beta =0$), we obtain an infinite
sequence of  concentric rings, with angular radius
\begin{equation}
\theta ^{E}_{n}=\left( 1-\frac {\zeta _{n}}{c_{3}}\right) \theta
^{0}_{n}, \label{pm16} 
\end{equation}
 known as the Einstein rings (ERs). The magnification of the images is given by
 $$\mu _{n}=\frac{1}{\beta}\frac{\theta
^{0}_{n}\zeta _{n}}{c_{3}}$$
As we see from eqn \ref{pm8} that $\zeta_n$ gets exponentially suppressed for increasing $n$
, the first image is the brightest and the brightness gets progressively decreased for increasing $n$.
In accordance with Bozza \cite{Bozza:2002zj} we consider a situation where one is able to resolve only the first relativistic image from the rest 
and compare the brightness of the first image with the cumulative brightness of all other relativistic images. For this purpose, we define
lensing observables
\begin{equation}
 s=\theta
_{1}-\theta _{\infty } \label{g1a},
\end{equation}
the separation between the outermost image and rest of the images;
and
\begin{equation}
 r=\frac{\mu _{1}}{\sum\limits_{n=2}^{\infty }\mu _{n}},
\label{g1b}
\end{equation}
the luminosity ratio of outermost image and rest of the images (clumped together);
\\
as well as 
\begin{equation}
\theta_{\infty}=u_{ps}/d_{ol},
\end{equation}
the angular location of black hole shadow.

For high alignment approximation the above equations get simplified to 
\begin{equation}
s=\theta_{\infty }e^{(c_{2}-2\pi )/c_{1}}, \label{g2a}
\end{equation} and
\begin{equation}
 r=e^{2\pi /c_{1}}+e^{c_{2}/c_{1}}-1,
\label{g2b} 
\end{equation}
 This gives a link between the calculated quantities $c_1,c_2,u_{ps}$ and observational quantities $r,s,\theta_{\infty}.$
%$$(c_1,c_2,u_{ps})\longleftrightarrow(r,s,\theta_{\infty}).$$

\end{appendices}


\begin{thebibliography}{27}
\expandafter\ifx\csname natexlab\endcsname\relax\def\natexlab#1{#1}\fi
\expandafter\ifx\csname bibnamefont\endcsname\relax
  \def\bibnamefont#1{#1}\fi
\expandafter\ifx\csname bibfnamefont\endcsname\relax
  \def\bibfnamefont#1{#1}\fi
\expandafter\ifx\csname citenamefont\endcsname\relax
  \def\citenamefont#1{#1}\fi
\expandafter\ifx\csname url\endcsname\relax
  \def\url#1{\texttt{#1}}\fi
\expandafter\ifx\csname urlprefix\endcsname\relax\def\urlprefix{URL }\fi
\providecommand{\bibinfo}[2]{#2}
\providecommand{\eprint}[2][]{\url{#2}}

\bibitem[{\citenamefont{Braunstein et~al.}(2013)\citenamefont{Braunstein,
  Pirandola, and Zyczkowski}}]{Braunstein:2009my}
\bibinfo{author}{\bibfnamefont{S.~L.} \bibnamefont{Braunstein}},
  \bibinfo{author}{\bibfnamefont{S.}~\bibnamefont{Pirandola}},
  \bibnamefont{and}
  \bibinfo{author}{\bibfnamefont{K.}~\bibnamefont{Zyczkowski}},
  \bibinfo{journal}{Phys.Rev.Lett.} \textbf{\bibinfo{volume}{110}},
  \bibinfo{pages}{101301} (\bibinfo{year}{2013}), \eprint{0907.1190}.

\bibitem[{\citenamefont{Almheiri et~al.}(2013)\citenamefont{Almheiri, Marolf,
  Polchinski, and Sully}}]{Almheiri:2012rt}
\bibinfo{author}{\bibfnamefont{A.}~\bibnamefont{Almheiri}},
  \bibinfo{author}{\bibfnamefont{D.}~\bibnamefont{Marolf}},
  \bibinfo{author}{\bibfnamefont{J.}~\bibnamefont{Polchinski}},
  \bibnamefont{and} \bibinfo{author}{\bibfnamefont{J.}~\bibnamefont{Sully}},
  \bibinfo{journal}{JHEP} \textbf{\bibinfo{volume}{1302}}, \bibinfo{pages}{062}
  (\bibinfo{year}{2013}), \eprint{1207.3123}.

\bibitem[{\citenamefont{Mathur}(2009)}]{Mathur:2009hf}
\bibinfo{author}{\bibfnamefont{S.~D.} \bibnamefont{Mathur}},
  \bibinfo{journal}{Class.Quant.Grav.} \textbf{\bibinfo{volume}{26}},
  \bibinfo{pages}{224001} (\bibinfo{year}{2009}), \eprint{0909.1038}.

\bibitem[{\citenamefont{Pen}(2013)}]{Pen:2013qva}
\bibinfo{author}{\bibfnamefont{U.-L.} \bibnamefont{Pen}}
  (\bibinfo{year}{2013}), \eprint{1312.4017}.

\bibitem[{\citenamefont{Bhadra}(2003)}]{Bhadra:2003zs}
\bibinfo{author}{\bibfnamefont{A.}~\bibnamefont{Bhadra}},
  \bibinfo{journal}{Phys.Rev.} \textbf{\bibinfo{volume}{D67}},
  \bibinfo{pages}{103009} (\bibinfo{year}{2003}), \eprint{gr-qc/0306016}.

\bibitem[{\citenamefont{Ding et~al.}(2011)\citenamefont{Ding, Kang, Chen, Chen,
  and Jing}}]{Ding:2010dc}
\bibinfo{author}{\bibfnamefont{C.}~\bibnamefont{Ding}},
  \bibinfo{author}{\bibfnamefont{S.}~\bibnamefont{Kang}},
  \bibinfo{author}{\bibfnamefont{C.-Y.} \bibnamefont{Chen}},
  \bibinfo{author}{\bibfnamefont{S.}~\bibnamefont{Chen}}, \bibnamefont{and}
  \bibinfo{author}{\bibfnamefont{J.}~\bibnamefont{Jing}},
  \bibinfo{journal}{Phys.Rev.} \textbf{\bibinfo{volume}{D83}},
  \bibinfo{pages}{084005} (\bibinfo{year}{2011}), \eprint{1012.1670}.

\bibitem[{\citenamefont{Stefanov et~al.}(2010)\citenamefont{Stefanov,
  Yazadjiev, and Gyulchev}}]{Stefanov:2010xz}
\bibinfo{author}{\bibfnamefont{I.~Z.} \bibnamefont{Stefanov}},
  \bibinfo{author}{\bibfnamefont{S.~S.} \bibnamefont{Yazadjiev}},
  \bibnamefont{and} \bibinfo{author}{\bibfnamefont{G.~G.}
  \bibnamefont{Gyulchev}}, \bibinfo{journal}{Phys.Rev.Lett.}
  \textbf{\bibinfo{volume}{104}}, \bibinfo{pages}{251103}
  (\bibinfo{year}{2010}), \eprint{1003.1609}.

\bibitem[{\citenamefont{Wei et~al.}(2011)\citenamefont{Wei, Liu, and
  Guo}}]{Wei:2011zw}
\bibinfo{author}{\bibfnamefont{S.-W.} \bibnamefont{Wei}},
  \bibinfo{author}{\bibfnamefont{Y.-X.} \bibnamefont{Liu}}, \bibnamefont{and}
  \bibinfo{author}{\bibfnamefont{H.}~\bibnamefont{Guo}},
  \bibinfo{journal}{Phys.Rev.} \textbf{\bibinfo{volume}{D84}},
  \bibinfo{pages}{041501} (\bibinfo{year}{2011}), \eprint{1103.3822}.

\bibitem[{\citenamefont{Modesto}(2008)}]{modesto1}
\bibinfo{author}{\bibfnamefont{L.}~\bibnamefont{Modesto}}
  (\bibinfo{year}{2008}), \eprint{0811.2196}.

\bibitem[{\citenamefont{Rovelli}(2007)}]{Loop}
\bibinfo{author}{\bibfnamefont{C.}~\bibnamefont{Rovelli}},
  \emph{\bibinfo{title}{Quantum Gravity}}, Cambridge Monographs on Mathematical
  Physics (\bibinfo{publisher}{Cambridge University Press},
  \bibinfo{year}{2007}), ISBN \bibinfo{isbn}{9780521715966},
  \urlprefix\url{http://books.google.co.in/books?id=THBBPwAACAAJ}.

\bibitem[{\citenamefont{Bilson-Thompson and
  Vaid}(2014)}]{Bilson-Thompson:2014hoa}
\bibinfo{author}{\bibfnamefont{S.}~\bibnamefont{Bilson-Thompson}}
  \bibnamefont{and} \bibinfo{author}{\bibfnamefont{D.}~\bibnamefont{Vaid}}
  (\bibinfo{year}{2014}), \eprint{1402.3586}.

\bibitem[{\citenamefont{Caravelli and Modesto}(2010)}]{modesto2}
\bibinfo{author}{\bibfnamefont{F.}~\bibnamefont{Caravelli}} \bibnamefont{and}
  \bibinfo{author}{\bibfnamefont{L.}~\bibnamefont{Modesto}},
  \bibinfo{journal}{Class.Quant.Grav.} \textbf{\bibinfo{volume}{27}},
  \bibinfo{pages}{245022} (\bibinfo{year}{2010}), \eprint{1006.0232}.

\bibitem[{\citenamefont{Modesto and Premont-Schwarz}(2009)}]{Modesto:2009ve}
\bibinfo{author}{\bibfnamefont{L.}~\bibnamefont{Modesto}} \bibnamefont{and}
  \bibinfo{author}{\bibfnamefont{I.}~\bibnamefont{Premont-Schwarz}},
  \bibinfo{journal}{Phys.Rev.} \textbf{\bibinfo{volume}{D80}},
  \bibinfo{pages}{064041} (\bibinfo{year}{2009}), \eprint{0905.3170}.

\bibitem[{\citenamefont{Keeton and Petters}(2005)}]{Keeton:2005jd}
\bibinfo{author}{\bibfnamefont{C.~R.} \bibnamefont{Keeton}} \bibnamefont{and}
  \bibinfo{author}{\bibfnamefont{A.}~\bibnamefont{Petters}},
  \bibinfo{journal}{Phys.Rev.} \textbf{\bibinfo{volume}{D72}},
  \bibinfo{pages}{104006} (\bibinfo{year}{2005}), \eprint{gr-qc/0511019}.

\bibitem[{\citenamefont{Shapiro et~al.}(2004)\citenamefont{Shapiro, Davis,
  Lebach, and Gregory}}]{PhysRevLett.92.121101}
\bibinfo{author}{\bibfnamefont{S.~S.} \bibnamefont{Shapiro}},
  \bibinfo{author}{\bibfnamefont{J.~L.} \bibnamefont{Davis}},
  \bibinfo{author}{\bibfnamefont{D.~E.} \bibnamefont{Lebach}},
  \bibnamefont{and} \bibinfo{author}{\bibfnamefont{J.~S.}
  \bibnamefont{Gregory}}, \bibinfo{journal}{Phys. Rev. Lett.}
  \textbf{\bibinfo{volume}{92}}, \bibinfo{pages}{121101}
  (\bibinfo{year}{2004}).

\bibitem[{\citenamefont{Lochan and Vaz}(2012)}]{Lochan:2012in}
\bibinfo{author}{\bibfnamefont{K.}~\bibnamefont{Lochan}} \bibnamefont{and}
  \bibinfo{author}{\bibfnamefont{C.}~\bibnamefont{Vaz}},
  \bibinfo{journal}{Phys.Rev.} \textbf{\bibinfo{volume}{D85}},
  \bibinfo{pages}{104041} (\bibinfo{year}{2012}), \eprint{1202.2301}.

\bibitem[{\citenamefont{Rovelli}(1996)}]{Rovelli:1996dv}
\bibinfo{author}{\bibfnamefont{C.}~\bibnamefont{Rovelli}},
  \bibinfo{journal}{Phys.Rev.Lett.} \textbf{\bibinfo{volume}{77}},
  \bibinfo{pages}{3288} (\bibinfo{year}{1996}), \eprint{gr-qc/9603063}.

\bibitem[{\citenamefont{{Darwin}}(1959)}]{dar59}
\bibinfo{author}{\bibfnamefont{C.}~\bibnamefont{{Darwin}}},
  \bibinfo{journal}{Royal Society of London Proceedings Series A}
  \textbf{\bibinfo{volume}{249}}, \bibinfo{pages}{180} (\bibinfo{year}{1959}).

\bibitem[{\citenamefont{{Darwin}}(1961)}]{dar61}
\bibinfo{author}{\bibfnamefont{C.}~\bibnamefont{{Darwin}}},
  \bibinfo{journal}{Royal Society of London Proceedings Series A}
  \textbf{\bibinfo{volume}{263}}, \bibinfo{pages}{39} (\bibinfo{year}{1961}).

\bibitem[{\citenamefont{Virbhadra and Ellis}(2000)}]{Virbhadra:2000ju}
\bibinfo{author}{\bibfnamefont{K.~S.} \bibnamefont{Virbhadra}}
  \bibnamefont{and} \bibinfo{author}{\bibfnamefont{G.~F.~R.}
  \bibnamefont{Ellis}}, \bibinfo{journal}{Phys. Rev. D}
  \textbf{\bibinfo{volume}{62}}, \bibinfo{pages}{084003}
  (\bibinfo{year}{2000}).

\bibitem[{\citenamefont{{Virbhadra} and {Ellis}}(2002)}]{nsl.ve}
\bibinfo{author}{\bibfnamefont{K.~S.} \bibnamefont{{Virbhadra}}}
  \bibnamefont{and} \bibinfo{author}{\bibfnamefont{G.~F.}
  \bibnamefont{{Ellis}}}, \bibinfo{journal}{\prd}
  \textbf{\bibinfo{volume}{65}}, \bibinfo{eid}{103004} (\bibinfo{year}{2002}).

\bibitem[{\citenamefont{Bozza}(2002)}]{Bozza:2002zj}
\bibinfo{author}{\bibfnamefont{V.}~\bibnamefont{Bozza}},
  \bibinfo{journal}{Phys.Rev.} \textbf{\bibinfo{volume}{D66}},
  \bibinfo{pages}{103001} (\bibinfo{year}{2002}), \eprint{gr-qc/0208075}.

\bibitem[{\citenamefont{Bozza and Mancini}(2004)}]{Bozza:2003cp}
\bibinfo{author}{\bibfnamefont{V.}~\bibnamefont{Bozza}} \bibnamefont{and}
  \bibinfo{author}{\bibfnamefont{L.}~\bibnamefont{Mancini}},
  \bibinfo{journal}{Gen.Rel.Grav.} \textbf{\bibinfo{volume}{36}},
  \bibinfo{pages}{435} (\bibinfo{year}{2004}), \eprint{gr-qc/0305007}.

\bibitem[{\citenamefont{Nicolini et~al.}(2006)\citenamefont{Nicolini,
  Smailagic, and Spallucci}}]{Nicolini:2005vd}
\bibinfo{author}{\bibfnamefont{P.}~\bibnamefont{Nicolini}},
  \bibinfo{author}{\bibfnamefont{A.}~\bibnamefont{Smailagic}},
  \bibnamefont{and}
  \bibinfo{author}{\bibfnamefont{E.}~\bibnamefont{Spallucci}},
  \bibinfo{journal}{Phys.Lett.} \textbf{\bibinfo{volume}{B632}},
  \bibinfo{pages}{547} (\bibinfo{year}{2006}), \eprint{gr-qc/0510112}.

\bibitem[{\citenamefont{Callan et~al.}(1989)\citenamefont{Callan, Myers, and
  Perry}}]{Callan:1988hs}
\bibinfo{author}{\bibfnamefont{C.~G}~\bibnamefont{Callan}}, \bibinfo{author}{\bibfnamefont{R.~C.}
  \bibnamefont{Myers}}, \bibnamefont{and}
  \bibinfo{author}{\bibfnamefont{M.~J.}~\bibnamefont{Perry}},
  \bibinfo{journal}{Nucl.Phys.} \textbf{\bibinfo{volume}{B311}},
  \bibinfo{pages}{673} (\bibinfo{year}{1989}).

\bibitem[{\citenamefont{Tsukamoto et~al.}(2012)\citenamefont{Tsukamoto, Harada,
  and Yajima}}]{Tsukamoto:2012xs}
\bibinfo{author}{\bibfnamefont{N.}~\bibnamefont{Tsukamoto}},
  \bibinfo{author}{\bibfnamefont{T.}~\bibnamefont{Harada}}, \bibnamefont{and}
  \bibinfo{author}{\bibfnamefont{K.}~\bibnamefont{Yajima}},
  \bibinfo{journal}{Phys.Rev.} \textbf{\bibinfo{volume}{D86}},
  \bibinfo{pages}{104062} (\bibinfo{year}{2012}), \eprint{1207.0047}.

\bibitem[{\citenamefont{Maggiore}(2008)}]{Maggiore:2007nq}
\bibinfo{author}{\bibfnamefont{M.}~\bibnamefont{Maggiore}},
  \bibinfo{journal}{Phys.Rev.Lett.} \textbf{\bibinfo{volume}{100}},
  \bibinfo{pages}{141301} (\bibinfo{year}{2008}), \eprint{0711.3145}.

\end{thebibliography}
\end{document}